\documentclass[doublespacing]{elsart}

\usepackage{graphicx}

\usepackage{amssymb}

\usepackage{rotating}

\begin{document}

\begin{frontmatter}


\title{GABRIELA\thanksref{funds} : a new detector array for $\gamma-$ray and
conversion electron spectroscopy of transfermium elements.}

\thanks[funds]{This project is jointly funded by the JINR and the IN2P3/CNRS.
Work at the FLNR was performed partially under the financial support of the
Russian Foundation for Basic Research, contract N 05-02-16198 and the
JINR - BMBF (Germany), JINR - Polish and JINR - Slovak Cooperation Programmes.}


\author[lab1]{K. Hauschild\corauthref{me}},
\corauth[me]{Corresponding author.}
\ead{hauschild@csnsm.in2p3.fr}
\author[lab2]{A.V. Yeremin},
\author[lab3]{O. Dorvaux},
\author[lab1]{A. Lopez-Martens},
\author[lab2]{A.V. Belozerov},
\author[lab1]{Ch. Brian\c{c}on},
\author[lab2]{M.L. Chelnokov},
\author[lab2]{V.I. Chepigin},
\author[lab1]{S.A. Garcia--Santamaria},
\author[lab2]{V.A. Gorshkov},
\author[lab4]{F. Hanappe},
\author[lab2]{A.P. Kabachenko},
\author[lab1]{A. Korichi}, 
\author[lab2]{O.N. Malyshev},
\author[lab2]{Yu. Ts. Oganessian},
\author[lab2]{A.G. Popeko},
\author[lab3]{N. Rowley},
\author[lab2]{A.V. Shutov},
\author[lab3]{L. Stuttg\'{e}},
\author[lab2]{A.I. Svirikhin}


\address[lab1]{CSNSM, IN2P3-CNRS, F-91405 Orsay Campus, France}
\address[lab2]{Flerov Laboratory of Nuclear Reactions, JINR, 141 980 Dubna,
  Russia}
\address[lab3]{IReS, IN2P3-CNRS, F-67037 Strasbourg, France}
\address[lab4]{Universit\'{e} Libre de Bruxelles, C.P. 229, B-1050 Bruxelles,
 Belgium}

\begin{abstract}
With the aid of the Geant4 Monte Carlo simulation package a new detection
system has been designed for the focal plane of the recoil separator
VASSILISSA situated at the Flerov Laboratory of Nuclear Reactions,
JINR, Dubna. \textbf{GABRIELA} (\textbf{G}amma \textbf{A}lpha \textbf{B}eta
\textbf{R}ecoil \textbf{I}nvestigations with the \textbf{E}lectromagnetic
\textbf{A}nalyser VASSILISSA) has been optimised to detect the arrival of
reaction products and their subsequent radioactive decays involving the
emission of $\alpha$-- and $\beta$--particles, fission fragments,
$\gamma$-- and X--rays and conversion electrons. The new detector system is
described and the results of the first commissioning experiments are presented.
\end{abstract}

\begin{keyword}
Recoil separators; Decay tagging spectrometer;
Alpha, gamma-ray, and conversion electron spectroscopy at recoil separators;
GEANT Monte Carlo simulations;

\PACS 
\sep 23.60.+e  
\sep 27.80.+w  
\sep 29.30.Dn  
\sep 29.30.Kv  

\end{keyword}
\end{frontmatter}

\clearpage


 \section{Introduction}
The heaviest elements provide a unique laboratory to study nuclear structure
and nuclear dynamics under the influence of large Coulomb forces and large
mass ($A$). 
%
%
The stability of nuclei beyond the spherical ``doubly--magic''
$^{208}$Pb ($Z=82,~N=126$) decreases rapidly until the transfermium region
($Z>100$) where a lowering of the level density of single-particle states
for nuclei in the neighbourhood of the deformed doubly--magic
$^{270}_{108}$Hs reverses this trend locally \cite{Pat1991}. However,
the position of the spherical doubly-magic nucleus beyond $^{208}$Pb remains
controversial : recent calculations predicting $Z=$ 114, 120, or, 126 for the
next magic proton shell, and $N=$ 172 or 184 for neutrons
\cite{Cwiok1996,Bender1999,Kruppa2000}.
Among other things, this is a consequence of the treatment
of the spin-orbit splitting. At large values of $A$ a weakening of the
spin-orbit splitting is predicted \cite{Bohr,Mairle} which results in the
lowering of orbitals with $l\!=\!N$, $j\!=\!l - \frac{1}{2}$.
The magnitude of this effect can either create or destroy stabilising gaps
in the single-particle spectrum. For example, in various models
the gap predicted at $Z=114$ depends highly on the $2f_{7/2}$ and
$2f_{5/2}$ proton spin--orbit splitting.
%
%
%
It is therefore crucial to determine the relative
excitation energies of these single-particle states in the
transfermium region \cite{Chas1997} to reduce the
extrapolation required in predicting the position of
this ``island of stability'' for the very heaviest nuclei \cite{Afan2003}.
Recent reviews can be found in Refs. \cite{Leino2004,Cwiok2005}.\\

Beyond Einsteinium (Z=99) detailed spectroscopic data is sparse. In both
$^{256}$Fm \cite{Hall1989} and $^{255}$Fm \cite{Ahmad2000} $\gamma-$ray
spectroscopy was performed after the chemical separation of
reaction products following the irradiation of $^{254}$Es and
$^{253}$Es targets respectively.
Unfortunately further studies using this method are hindered by a lack of
suitable targets. Another method to populate the nuclei of interest is
via heavy--ion fusion evaporation (HI,xn) reactions. In this case
it is the overwhelming background from the predominant fission channel
that needs to be addressed. This has been achieved
with gas--jet transport systems and in-flight recoil separators.
%
%
Recently spectroscopic studies in this mass region have seen intense
activity in two distinct directions : 1) prompt in--beam spectroscopy
at the target position exploiting the recoil decay tagging (RDT)
method and, 2) isomeric, and, or, decay spectroscopy at the focal plane
of the recoil separator. A number of rotational bands have now been observed
using both $\gamma-$ray and conversion electron (CE) spectroscopy :
$^{254}$No \cite{Reiter1999,Leino1999,Butler2002,Humphreys2004},
$^{252}$No \cite{Herzberg2001},
$^{253}$No \cite{Butler2002,Reiter2005},
$^{250}$Fm \cite{Herzberg2003},
$^{251}$Md \cite{Chatillon2003} and $^{255}$Lr \cite{Greenlees2005}.
These results and additional unpublished data
have been reviewed in \cite{Herz2004}.
However, focal plane decay studies using $\alpha-\gamma$ coincidence
measurements have only been reported for a few transfermium nuclei :
$^{251}$No \cite{Hess2004},
$^{253}$No \cite{Herz2004,Hess2003} and $^{255}$Rf \cite{Hess2001}.
The $\alpha-\gamma$ coincidence, and, $\alpha-$CE  coincidence
decay spectroscopy of $^{257}$No \cite{Asai2005} presents
an interesting development with the re-emergence of gas-jet systems.

In these high Z nuclei the internal conversion becomes an
extremely important decay mode since it can compete effectively with
gamma decay. This makes it essential to perform electron spectroscopy
and is the motivation behind the projects GREAT \cite{GREAT} and
BEST, and, the subject of this paper, \textbf{GABRIELA}.
In section 2 the salient features of the VASSILISSA set up will
be presented. Then, in section 3, the modifications to the experimental
set-up needed to perform detailed spectroscopy of excited states in
transfermium nuclei are described along with the electronics developments
required for the programme. Finally, some experimental results from
commissioning runs will be shown 
to illustrate the performance of the GABRIELA system.

\section{The VASSILISSA separator system}
\label{vass}
In the following section only a brief description of the VASSILISSA separator
will be given. More details can be found in Ref \cite{Yer1997,Mal2000}.\\
The principal component of VASSILISSA consists of three electrostatic
dipoles which separate spatially the trajectories of the recoiling nuclei,
multinucleon transfer reaction products, fission fragments and beam particles
by virtue of differences in their energies and ionic charges. An additional
dipole magnet deflects the evaporation residues (ER's) by 37$^\circ$ improving
the background suppression of the scattered beam by a factor of 10 - 50. This
magnet also acts as a mass analyser \cite{Pop2003,Mal2004}.
Between the magnet and the separator system there is a 2 m thick concrete
wall which provides substantial
shielding from the beam dump. Downstream of the magnet a time-of-flight
measurement is made and the ER's are then implanted into a 300 $\mu$m thick,
16--strip, 58 $\times$ 58 mm$^2$ position sensitive Si detector at the focal
plane of the separator (hereafter called the stop detector).
Each strip is position sensitive in the vertical direction
with a resolution of 0.3 - 0.5 mm (obtained from $\alpha-\alpha$ correlations)
and has a typical energy resolution of 20 keV for 5-10 MeV alpha particles.
The subsequent position- and time--correlated alpha decays, characteristic of
the implanted recoils, are also measured in the Si detector. The detection
efficiency for these $\alpha$ particles is around 50$\%$.\\

\section{GABRIELA}
\label{gabriela}
In order to perform gamma--ray and conversion--electron spectroscopy at the
focal plane of VASSILISSA a number of modifications were needed. The Monte
Carlo simulation code Geant4 \cite{geant4} has been used as an aid to design
an experimental set up with the goal of maximising the efficiency and
resolution with a minimum of complexity. The set up is given in more
details in the following subsections.

\subsection{Stop detector and support}
\begin{figure}
\begin{center}
\includegraphics[width=13.5cm]{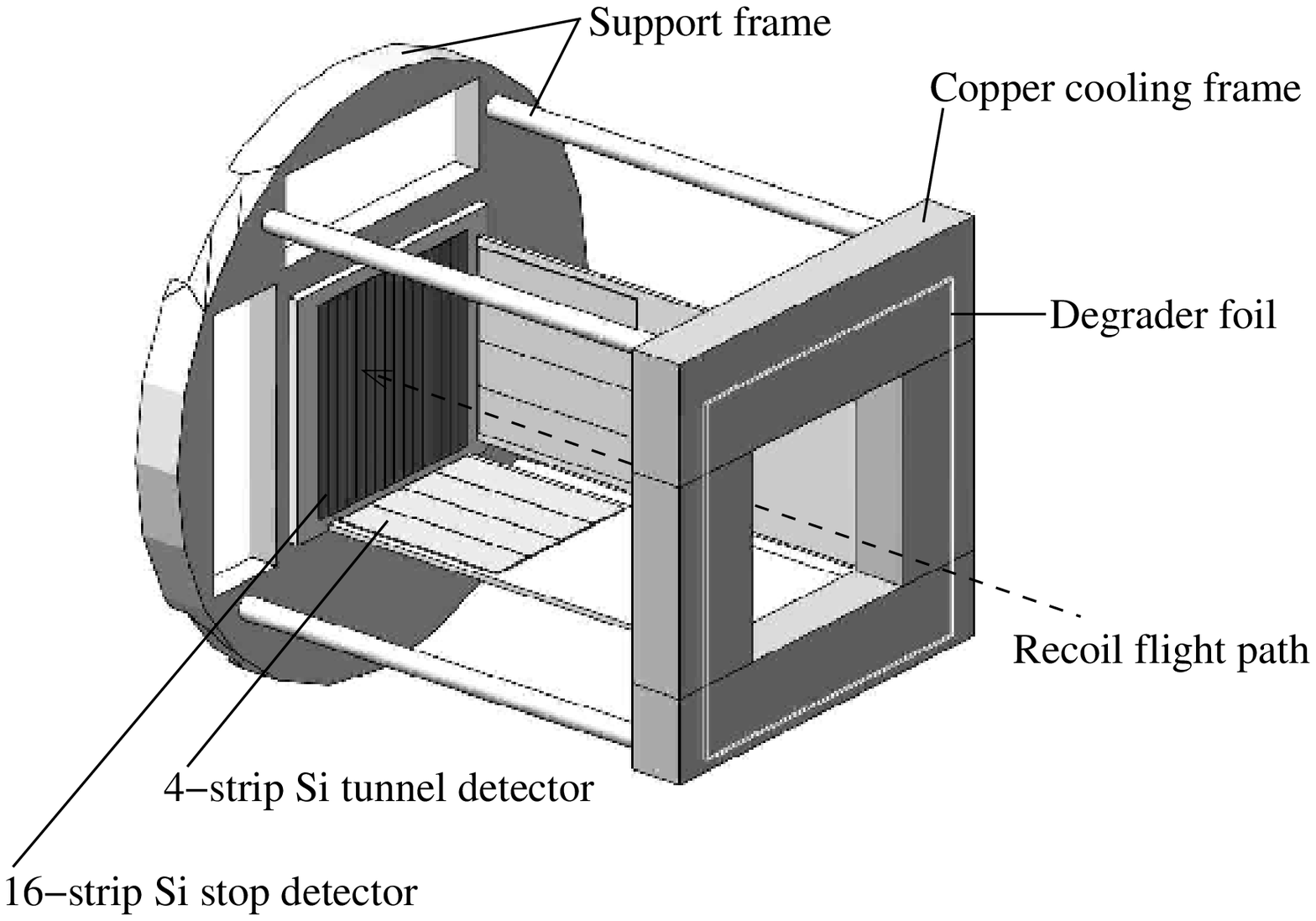}
\end{center}
\caption{Schematic view of the Si stop and tunnel detector
set-up at the focal plane of VASSILISSA. Two sides of the tunnel have
been removed for clarity.
}
\label{tunnel}
\end{figure}

A new detector system, including a new more compact vacuum chamber optimised
for transparency to $\gamma$--rays, was constructed to replace the old system
which was used to measure alpha decay and spontaneous fission.
The new aluminium chamber has a of thickness of 6.5 mm, with the portion 
in front of the Ge detectors machined down to 2.5 mm,
and an inner diameter of 160 mm. The support
for the stop detector has been made from a single disc of
stainless steel with cut-outs to allow cable connectors, cooling
fluid feed--through, and, more importantly, an unobstructed view of the
detector from the sides and from upstream. Fig. \ref{tunnel} shows a
schematic view.


\subsection{Germanium detector array for $\gamma$-ray spectroscopy}
\begin{figure}
\begin{center}
\includegraphics[width=13.5cm]{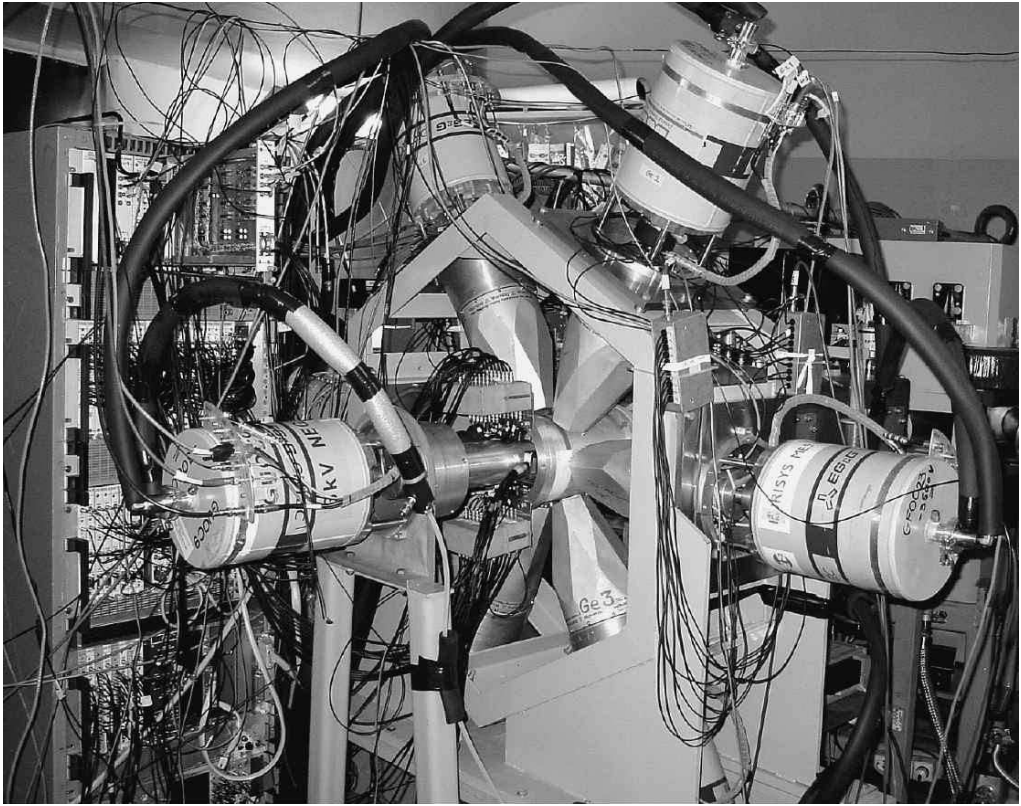}
\end{center}
\caption{Photograph of the GABRIELA set-up at the focal plane of VASSILISSA.
}
\label{gabriela_photo}
\end{figure}

The focal plane stop detector was surrounded by 7 Eurogam Phase-I Ge
detectors \cite{Bea92} obtained from the French-UK loan pool. Six of
these were placed inside BGO Compton shields and formed a ring around the
detector chamber with a focal point on the upstream (backwards) side of
the stop detector (see Fig. \ref{gabriela_photo}).
The distance from the centre of the stop detector to the front face of
Ge crystal for these six detectors was about 130 mm.
The aluminium back plate that closes the vacuum chamber was designed with 
an inset to enable the seventh Ge detector to be placed as close as
possible to stop detector (about 35 mm). In the hollowed out portion
the back plate is only 1.5 mm thick. The suppression shields served two
purposes. The first of which is to improve the peak-to-total by vetoing
events for which a $\gamma$ ray Compton scatters out of the Ge detector
which is indispensable for the identification of weak lines which would
have otherwise been buried under the Compton background of more intense lines.
The second is to reduce the counting rate from background radiation by
vetoing events for which $\gamma$ rays emitted from the concrete walls
(mainly $^{40}$K) interact in the germanium detectors. This enables
increased ``search'' times to be used in the hunt for long lived isomers.

\begin{figure}
\begin{center}
\includegraphics[width=13.5cm]{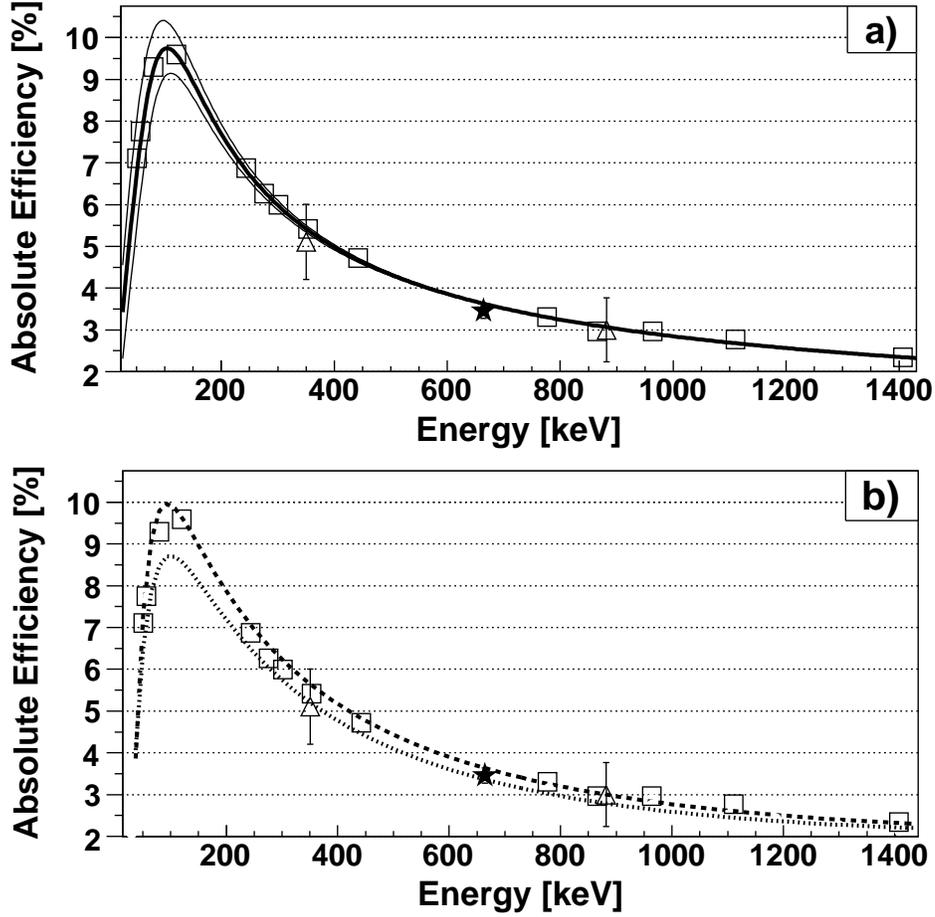}
\end{center}
\caption{Absolute efficiency curve for the 7 Ge detectors.
$\blacksquare$ : measured using $^{133}$Ba, $^{152}$Eu and $^{241}$Am sources;
 $\star$ measured using $\gamma$--electron
coincidences from the decay of an isomeric state in $^{207}$Rn implanted into
the stop detector using the reaction $^{164}$Dy($^{48}$Ca,$5n$); $\triangle$
measured using $\alpha-\gamma$ coincidences from the fine--structure
decay of $^{211}$Bi [$E_\gamma=351$-keV] and $^{217}$Th [$E_\gamma=882$-keV].
a) the bold solid line : fit to the data using the expression
$log(\epsilon) = [(A + Bx_1 + Cx_1^2)^{-G} + (D + Ex_2 + Fx_2^2)^{-G}]^{-1/G}$
where $x_1 = E_\gamma/100$ and $x_2 = E_\gamma/1000$; the thin solid lines
represent the error in the fit. b) dashed line : Geant4 simulated
efficiency curve for a point source positioned at the centre of the
stop detector; dotted line : Geant4 simulated efficiency curve for
a distributed source.
}
\label{gam_eff}
\end{figure}

To obtain an absolute efficiency curve for $\gamma$-ray detection
$^{133}$Ba, $^{152}$Eu and $^{241}$Am sources of known activities were
attached individually to the centre of an old stop detector
which was then fixed to the detector support and inserted into the chamber.
This permitted calibrations to be taken in conditions as close as possible
to those during experimental runs. The sole difference being that the
calibrations were performed with a point source, while experimental
data is taken with the gamma-ray emitting recoils distributed almost
uniformly over the surface of the stop detector. In Fig. \ref{gam_eff}a
the measured $\gamma$--ray photo-peak efficiency from the calibration data is
presented.

The reaction $^{174}$Yb($^{48}$Ca,$xn$)$^{222-x}$Th,
which is used primarily for alpha calibration purposes, can provide
$\gamma$ ray detection efficiency data under experimental conditions.
$^{217}$Th $\alpha$ decays to $^{213}$Ra. A fraction of these decays
populate excited states in $^{213}$Ra which, subsequently, decay to the
ground state via $\gamma$ emission \cite{Hess2002}. From a comparison of
the prompt $\alpha-\gamma$ coincidence intensity ($I(\alpha-\gamma)$)
with the total $\alpha$ singles spectrum ($I(\alpha)$) one can determine the
$\gamma$ ray efficiency after correcting for internal conversion
($\alpha_{TOT}$). That is :
$\epsilon_\gamma = I(\alpha-\gamma)/I(\alpha)\times(1~+~\alpha_{TOT})$.
This measurement has also been carried out for the transfer product
$^{211}$Bi which has fine--structure alpha decay to an excited state in
$^{207}$Tl. These data, represented by the $\triangle$ symbol in
Fig.\ref{gam_eff}a, are in agreement with the source data within errors.
However, they do appear to indicate an experimental detectection efficiency
lower than that taken with calibration sources.

To examine the effect a distributed source has on the efficiency,
Geant4 simulations have been performed for various $\gamma$-ray energies with
$\gamma$ rays emitted into 4$\pi$ from 1) a fixed point at the centre of the
stop detector, and, 2) a uniform distribution over the x-y plane of
the stop detector. The deposited energy recorded in the simulations was
taken from the secondary electrons which are created by the Compton
scattering and photoelectric processes. The results of these simulations
are presented in Fig. \ref{gam_eff}b and indicate that an energy dependent
scaling factor of 0.85 -- 0.96 is needed to map the point source efficiency onto
the distributed source efficiency. In order to obtain the excellent agreement
between the measured and simulated efficiency curves every germanium crystal
was shifted by 5 mm backwards within its aluminium housing relative to the
nominal values given in Daresbury technical drawing A0-36/8813.
This minor discrepancy is not that alarming since the precise dimensions
and positions for the individual germanium crystals are not known.

For the above Germanium detector measurements new spectroscopy
amplifiers and ADCs (4096 channels,
2 $\mu$s conversion time) which accept a veto signal were developed
at the Flerov Laboratory. They demonstrated an
excellent stability : during 1 month of measurements the omnipresent 1461 keV
background line from $^{40}$K was observed to have a maximum energy shift of
$< \pm 0.03 \%$ (0.4 keV).
Within the energy range of 81 - 1408 keV the rms deviation
of measured $\gamma$--ray energies compared to standard values \cite{TOI}
was found to be 0.1-, 0.2-, 0.2-, 0.2-, 0.2-, 0.2- and 0.1 keV
for the 7 Ge detectors and indicates the precision to be expected
in subsequent measurements. At 1332 keV a full width at half maximum (FWHM)
of $\sim$2.5 keV was obtained.\\


\subsection{Silicon detector array for conversion--electron  spectroscopy}
In the backward direction of the stop detector an array of four
4-strip silicon detectors (Canberra PF-4CT-50*50-500RM)
are arranged in a tunnel configuration
which is shown schematically in Fig. \ref{tunnel}.
Each detector has a total active area of 50 $\times$ 50 mm$^2$, a
thickness of 500 $\mu$m, a front-face dead-layer thickness of $<$25 nm Si
equivalent, and is mounted on a 1.6 mm thick, 60 $\times$
120 mm$^2$ IS450 resin board manufactured by ISOLA.
The pre-amplifiers (designed by GANIL) for each strip are mounted on the
reverse side of this support board which is attached to
a copper frame through which cooling fluid can be circulated. Thus the
heating effect of the pre-amplifiers can be counteracted and the Si detectors
cooled in order to reduce the resolution destroying leakage current.

The Si detectors are used to measure
emitted particles escaping from the stop detector : principally
conversion electrons, but also, alphas, fission fragments and betas.
Due to the tunnel geometry a large proportion of these particles will be
detected at small distances upstream from the stop detector.
The particles detected in the tunnel close to the stop detector
will have suffered more straggling than those detected
further upstream because, on average, they have travelled
further in the stop detector. Therefore, Geant4 simulations were performed
with the tunnel detectors being placed upstream from the stop detector at
various distances to evaluate the compromise
 between the geometrical detection efficiency and spectral resolution.
Dead layer thicknesses, support frames and epoxy boards
for the silicon detectors and the vacuum chamber were included in the
simulation geometry. Electrons were emitted into 4$\pi$
from an implantation depth of 3.0(5) $\mu$m distributed uniformly over
the x-y plane of the stop detector. The results of these simulations indicate
that the effect of straggling has a much smaller effect on the energy
resolutions compared to the degredation expected from the leakage current
in the detectors and the electronics noise in the system.
We have therefore tried to minimise the distance
between the stop and tunnel detectors with the nominal distance
between the epoxy support boards being about 2 mm.

\begin{figure}
\begin{center}
\includegraphics[width=13.5cm]{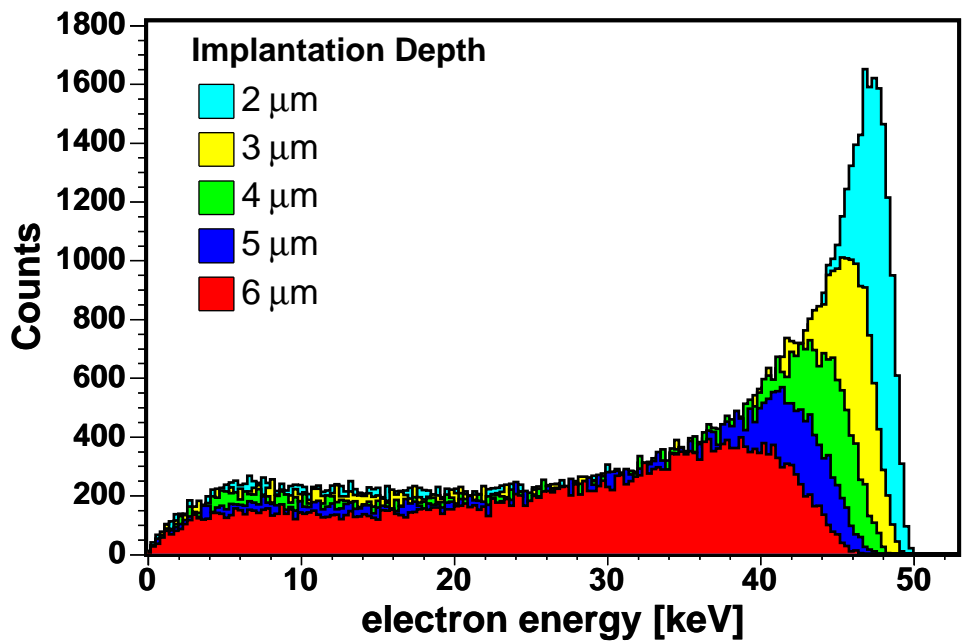}
\end{center}
\caption{Simulated energy deposited in the tunnel detectors for a 50 keV
electron emitted from varying implantation depths within the stop
detector.}
\label{50keV}
\end{figure}

Simulations have also been performed to investigate the
efficiency and resolution as a function of recoil implantation depth.
Using the geometry described above, with a 2 mm gap between the tunnel and
stop detector supports, electrons were simulated to have been emitted
from implantation depths of 2.0(5), 3.0(5), 4.0(5), 5.0(5) and 6.0(5)
$\mu$m. The results are given in Table \ref{electrons2} and can be broken into
3 regions : 1) For electron energies above 500 keV the performance of the set
up is almost independent of the implantation depth for those depths
simulated. 2) For energies
between 100 and 400 keV there is a marginal difference in efficiency and a
noticeable increase in FWHM with increasing implantation depth.
3) Below 100 keV there is a significant degradation in both the detection
efficiency and the resolution.
In some simulations the effect of straggling was so large that peaks were
no longer discernible and indicates that
there is a limit below which one
cannot perform electron spectroscopy in the tunnel detectors.
This effect can be clearly seen in the simulations for 50 keV electrons shown
in Fig. \ref{50keV}. The key to reducing this lower limit as
far as possible is by placing a degrader foil in front of the stop detector
in order to reduce the implantation depth.
%

Another important effect visible in Fig. \ref{50keV}, and presented in a more
systematic manner in Table \ref{electrons2}, is the shift in electron energy
measured in the tunnel detectors. An energy calibration of the tunnel
detectors must account for this shift. This can be achieved by either
correcting unsealed source calibrations for the shifts given in
Table \ref{electrons2}, or, by implanting into the stop detector recoils
known to decay via conversion electron emission and performing an in-beam
calibration. 

Before performing in-beam experiments
initial calibrations with a $^{133}$Ba source were
performed to align the Si electronics channels.
Operating the Si detectors at $-5^\circ$C energy resolutions of between
8 - 10 keV FWHM were obtained for the 322-keV line, in-line
with expectations when noise and leakage current affects are taken into
account.

\section{Commissioning experiments}

To test the new detectors and electronics a series of commissioning
experiments were performed using the complete fusions reactions
$^{164}$Dy($^{48}$Ca,$xn$)$^{212-x}$Rn,
$^{174}$Yb($^{40}$Ar,$xn$)$^{214-x}$Ra and
$^{181}$Ta($^{40}$Ar,$xn$)$^{221-x}$Pa at beam energies corresponding
to the evaporation of 4 and 5 neutrons. 

\begin{figure}
\begin{center}
\includegraphics[width=13.5cm]{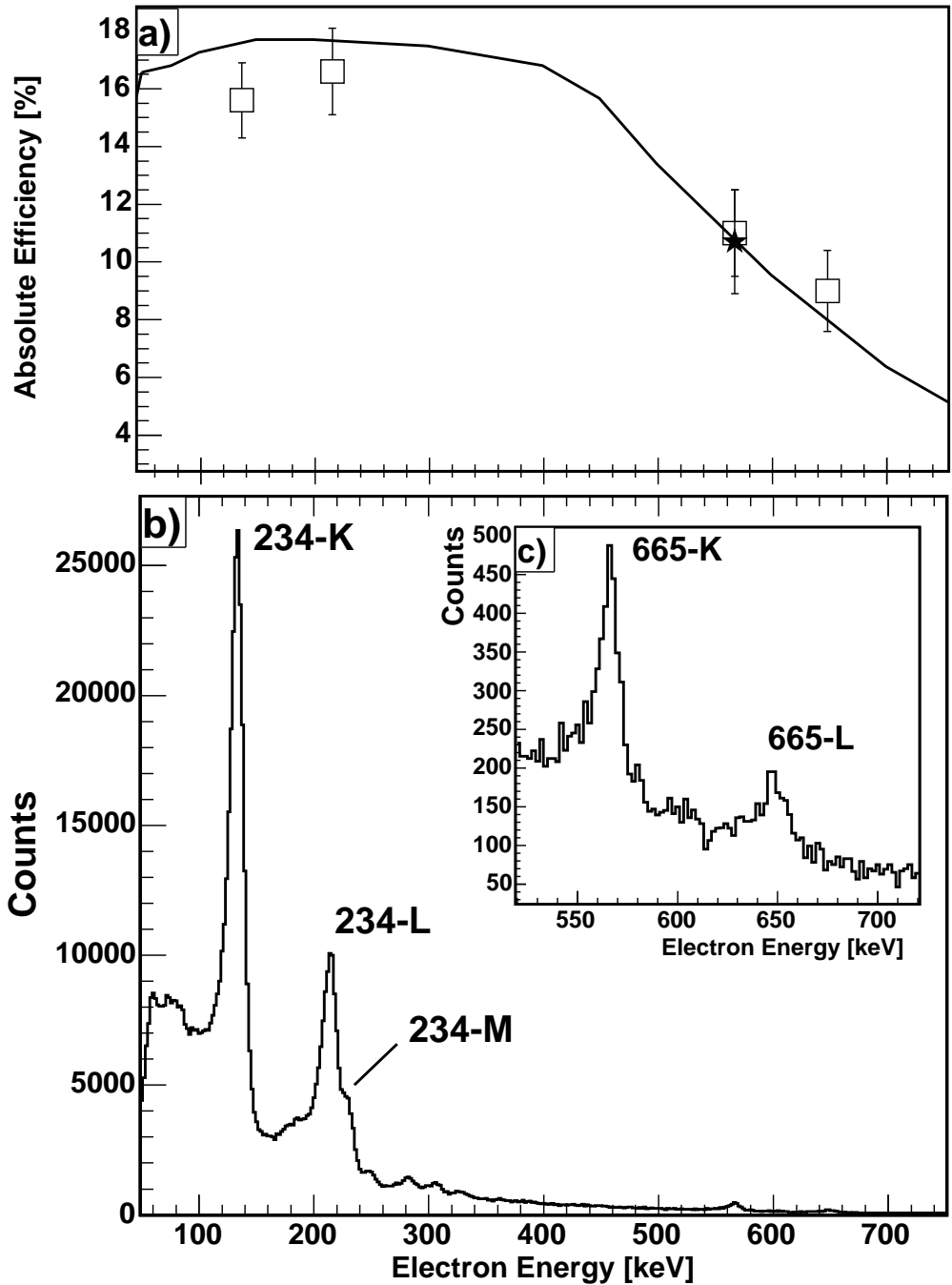}
\end{center}
\caption{a) Absolute efficiency curve for the 4 Si conversion electron
detectors. $\star$ : efficiency for the 567-keV (665 K conversion) obtained
from $\gamma$--conversion electron coincidence measurements following the
decay of the $13/2^+$ isomer in $^{207}$Rn; $\square$ : singles efficiency
measurements form the same $^{207}$Rn data; the bold solid line : results of
Geant4 simulations with an implantation depth of 3.0(5) $\mu$m distributed
uniformly in the x-y plane of the stop detector. b) Electron singles spectra
measured in the tunnel detectors within the time range
32 $\mu$s $<$ dT $>$ 1024 $\mu$s of a recoil from the reaction
$^{48}$Ca + $^{164}$Dy  being detected in the stop detector. c) The inset
shows the high energy part of the spectrum.
}
\label{electron_eff}
\end{figure}

The first reaction was used to obtain an absolute efficiency measurement for
the tunnel detectors.
In $^{207}$Rn a $13/2^+$ isomer at an excitation energy
of 899 keV decays with a half-life of 181(18) $\mu$s to an intermediate
$9/2^-$ state at 665 keV which then decays to the $5/2^-$ ground state
\cite{Rez74}. This results in a
$\sigma L:E$ = M2:234-keV\footnote{Where $\sigma$ represents either electric
or magnetic radiation and $L$ is the multipolarity and E is the transition
energy.}
transition followed by an E2:665-keV line. Using coincidence measurements and
singles intensities the following absolute efficiencies can be obtained :\\
1) $\epsilon_\gamma(665)~=~I(234e^-\otimes665\gamma)\times[1~+~\alpha_{TOT}^{E2}(665)]/I^{singles}_{e^-}(234)$, and,\\
2) $\epsilon_{e^-}(665-K)~=~I(234\gamma\otimes665-Ke^-)\times[1~+1/\alpha_K^{E2}(665)]/I^{singles}_{\gamma}(234)$, where
$I(E1 e^- \otimes E2 \gamma)$ represents the intensity observed for the
coincidence measurement between an electron of energy $E1$ and a
$\gamma$--ray of $E2$ and $\alpha$ is the conversion coefficient. To reduce
the possible contamination from other reaction channels all intensity
measurements were taken in a time range 32$\mu$s $<$ dT $<$ 1024$\mu$s with
respect to the implantation of a recoil in the stop detector. The good
agreement between the germanium array efficiency obtained
from the $e^--\gamma$ coincidence, $\epsilon_\gamma(665)$, and the source
measurements shown in Fig. \ref{gam_eff}a gives us confidence in the absolute
efficiency determined for the tunnel detector which have also been obtained
using the relationship :
$\epsilon_{e^-}(E:X)~=~I^{singles}_{e^-}(E:X)\times[1~+~1/\alpha_X(E)]/N_{isomer}$,
where, E is either 234- or 665-keV, X is either the K,L or M conversion
electron and
$N_{isomer}~=~I^{singles}_{\gamma}(665) \times \alpha_{TOT}^{E2}(665)/ \epsilon_\gamma(665)$. These absolute efficiencies are presented in Fig. \ref{electron_eff}a
and are in good agreement with the Geant4 simulations.
In Fig. \ref{electron_eff}b electron singles spectra are shown as an illustration of the quality of these data. The FWHM of the 234-K line ranges from 9.2-
to 16.1 keV depending on the strip with most of the strips having a FWHM of
10 - 11 keV.  The broadening of the lines with respect to the values
quoted in Table \ref{electrons2} can be attributed to the detectors not being
optimally cooled, an increase in noise during the U400 cyclotron operation
and an implantation depth of $>4\mu$m for these ``low'' Z recoils since
the thickness of the degrader used was optimised for the higher Z transfermiums.

As an aside, these data have allowed a more accurate measurement of the
half-life of the $13/2^+$ isomer in $^{207}$Rn. $\tau_{1/2} = 184.5(9) \mu s$ was
obtained for the 234-K conversion-electron transition using the method
described in Ref \cite{Leino81}.

\begin{figure}
\begin{center}
\includegraphics[width=13.5cm]{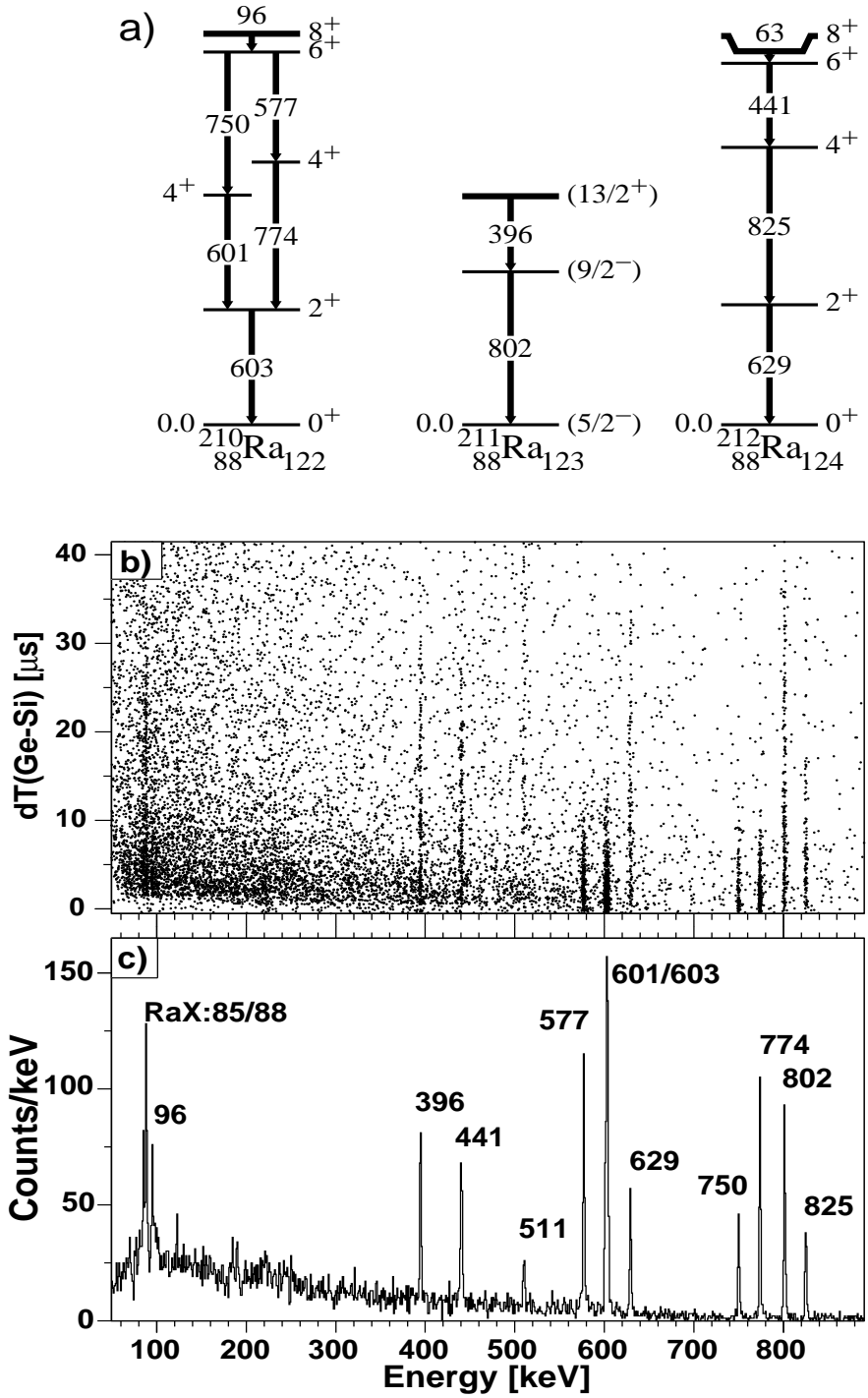}
\end{center}
\caption{a) Partial level schemes (taken from Refs.
\cite{Cocks,Res04,Hess04,Koh86})
below the isomeric states in the Radium isotopes
populated via the $^{174}$Yb($^{40}$Ar,$xn$)$^{214-x}$Ra reaction and
transported to the focal plane of VASSILISSA. b) The time difference between
recoil and $\gamma$-ray detection as a function of measured $\gamma$-ray
energy is plotted on an event-by-event basis. c) Delayed $\gamma$-ray
transitions observed within 40 $\mu$s after the recoil was detected.}
\label{Ra}
\end{figure}

$\gamma$-ray spectra obtained for an unsuppressed detector during the
$^{40}$Ar + $^{174}$Yb $\rightarrow$ $^{214}$Ra$^*$ test run are shown
in Fig. \ref{Ra}. The time difference between
recoil and $\gamma$-ray detection as a function of measured $\gamma$-ray
energy is plotted on an event-by-event basis in Fig. \ref{Ra}b. Different
lifetimes for transitions depopulating the different isomeric states are
clearly visible. $\gamma$--ray transitions detected within 40 $\mu$s
after the recoil implantation are shown in Fig. \ref{Ra}c.
In Table \ref{half-life} the
apparent half-lives of these $\gamma$--rays measured in the current work
are given. 
In view of the agreement obtained for the half-lives of
$8^+$ isomers in $^{210}$Ra and $^{212}$Ra compared to the published values
(2.28(8) $\mu$s cf 2.24, 2.1(1), and 2.36(4) $\mu$s \cite{Cocks,Res04,Hess04},
and 9.1(7) $\mu$s cf 10.5, 10.9(4) \cite{Hess04,Koh86}, respectively) the
discrepancy between our measured value of 9.7(6) $\mu$s and the value reported
in Ref. \cite{Hess04} of 4.0(5) $\mu$s
for the $(13/2^+)$ isomer in $^{211}$Ra needs independent confirmation.

\section{Conclusion and Perspectives}

The characteristics of the new GABRIELA detector array have been presented.
This array, designed with the aid Geant4 simulations, has been installed
at the focal plane of the VASSILISSA separator at the FLNR in Dubna. It has
been constructed with the goal of performing detailed spectroscopic studies
in transfermium nuclei. Following commissioning tests in May and June of 2004,
two one--month--long experimental campaigns were performed in
September - October 2004 and October 2005.
The complete fusion reactions $^{48}$Ca + $^{207,208}$Pb
$\rightarrow$ $^{255,256}$No$^*$ and $^{48}$Ca + $^{209}$Bi $\rightarrow$
$^{255}$Lr$^*$ were investigated. The decays of the isotopes
$^{253-255}$No, $^{255}$Lr and their daughter products are currently being
analysed.


\clearpage

\clearpage

\clearpage

\clearpage

\clearpage

\clearpage

\clearpage

\begin{table}
\caption{Results of Geant4 simulations with different ``recoil''
implantation depths in the stop detector.
For each simulation $2 \times 10^5$ electrons were emitted into 4$\pi$
from points distributed uniformly in the x-y plane of the stop detector.
The depth in the stop detector at which the electrons were emitted was
taken to be a Gaussian distribution with a sigma of 0.5 $\mu$m. $\Delta$E
is the difference between the emitted electron energy and the energy
measured in a tunnel detector (ie the energy deposited in the stop detector).
The efficiency, $\epsilon$, was determined
by integrating the simulated spectrum between $\pm3\sigma$ of the full
energy centroid. The error bars on $\epsilon$ are $< 0.2\%$ and arise from
the error in determining the integration limits.
The effect of electronics noise is not included.
}
\label{electrons2}
\tiny{
\begin{tabular}{|r||r|r|r||r|r|r||r|r|r||r|r|r|} \hline
         & \multicolumn{12}{|c|}{electron energy}\\
         & \multicolumn{3}{c||}{ 30 keV}
         & \multicolumn{3}{c||}{ 40 keV}
         & \multicolumn{3}{c||}{ 50 keV}
         & \multicolumn{3}{c|}{  75 keV} \\
depth    & $\Delta$E & fwhm     & $\epsilon$ & $\Delta$E & fwhm & $\epsilon$
         & $\Delta$E & fwhm     & $\epsilon$ & $\Delta$E & fwhm & $\epsilon$\\
($\mu$m) & (keV)     & (keV)    &   $\%$     & (keV)     & (keV)&   $\%$
         & (keV)     & (keV)    &   $\%$     & (keV)     & (keV)&   $\%$\\
\hline
 2.0(5)  & 4.6 & 8.2(1.3) & 13.8   
         & 3.2 & 6.3(0.5) & 16.4   
         & 2.2 & 4.1(1)   & 16.8   
         & 1.5 & 2.7(2)   & 17.0\\ 
 3.0(5)  & 7.8 & 9.1(2.2) &  7.6   
         & 5.3 & 8.3(1.1) & 14.8   
         & 3.8 & 6.2(3)   & 16.5   
         & 2.6 & 3.9(3)   & 16.8\\ 
 4.0(5)  &  -- &  --      &  --    
         & 8.0 & 9.3(0.2) & 11.2   
         & 5.9 & 8.8(6)   & 15.8   
         & 3.5 & 5.4(5)   & 16.7\\ 
 5.0(5)  & --  &  --      &  --    
         &11.1 &  12(2)   &  8.9   
         & 8.1 &  9.1(8)  & 13.1   
         & 4.7 &  6.6(2)  & 16.2\\ 
 6.0(5)  & --  &  --      &  --    
         & --  &  --      &  --    
         &10.7 & 13.6(2)  & 10.2   
         & 6.0 & 8.2(8)   & 15.1\\ 
\hline
         & \multicolumn{3}{c||}{100 keV} 
         & \multicolumn{3}{c||}{200 keV}  
         & \multicolumn{3}{c||}{300 keV}
         & \multicolumn{3}{c|}{ 400 keV}\\
depth    & $\Delta$E & fwhm     & $\epsilon$ & $\Delta$E & fwhm & $\epsilon$
         & $\Delta$E & fwhm     & $\epsilon$ & $\Delta$E & fwhm & $\epsilon$\\
($\mu$m) & (keV)     & (keV)    &   $\%$     & (keV)     & (keV)&   $\%$
         & (keV)     & (keV)    &   $\%$     & (keV)     & (keV)&   $\%$\\
\hline
 2.0(5)  & 1.2 & 2.2(1)   & 17.3   
         & 0.7 & 1.4(1)   & 17.5   
         & 0.5 & 1.2(1)   & 17.6   
         & 0.6 & 1.1(1)   & 16.9\\ 
 3.0(5)  & 2.0 & 3.4(1)   & 17.1   
         & 1.3 & 2.3(1)   & 17.5   
         & 1.0 & 2.0(1)   & 17.3   
         & 0.9 & 1.7(1)   & 16.8\\ 
 4.0(5)  & 2.8 & 4.3(4)   & 16.6   
         & 1.8 & 3.0(1)   & 17.3   
         & 1.5 & 2.5(2)   & 17.1   
         & 1.3 & 2.2(2)   & 16.5\\ 
 5.0(5)  & 3.6 & 5.1(3)   & 16.3   
         & 2.3 & 3.5(3)   & 16.8   
         & 1.8 & 3.0(2)   & 16.8   
         & 1.7 & 2.7(2)   & 16.4\\ 
 6.0(5)  & 4.5 & 6.0(6)   & 16.0   
         & 2.8 & 4.0(3)   & 16.3   
         & 2.3 & 3.3(3)   & 16.3   
         & 2.0 & 3.2(2)   & 16.2\\ 

\hline
         & \multicolumn{3}{c||}{500 keV} 
         & \multicolumn{3}{c||}{600 keV}  
         & \multicolumn{3}{c||}{700 keV}
         & \multicolumn{3}{c|}{ 800 keV}\\
depth    & $\Delta$E & fwhm     & $\epsilon$ & $\Delta$E & fwhm & $\epsilon$
         & $\Delta$E & fwhm     & $\epsilon$ & $\Delta$E & fwhm & $\epsilon$\\
($\mu$m) & (keV)     & (keV)    &   $\%$     & (keV)     & (keV)&   $\%$
         & (keV)     & (keV)    &   $\%$     & (keV)     & (keV)&   $\%$\\
 2.0(5)  & 0.4 & 1.1(1)  & 13.8   
         & 0.5 & 1.0(1)  &  9.5   
         & 0.5 & 1.0(1)  &  6.3   
         & 0.4 & 0.9(1)  &  4.2\\ 
 3.0(5)  & 0.8 & 1.6(1)  & 13.4
         & 0.8 & 1.5(1)  &  9.5
         & 0.7 & 1.4(2)  &  6.2
         & 0.7 & 1.3(2)  &  4.1\\
 4.0(5)  & 1.2 & 2.1(2)  & 13.4
         & 1.1 & 1.9(2)  &  9.3
         & 1.0 & 1.9(2)  &  6.2
         & 1.0 & 1.7(2)  &  4.1\\
 5.0(5)  & 1.5 & 2.5(2)  & 13.4
         & 1.4 & 2.4(1)  &  9.3
         & 1.4 & 2.3(2)  &  6.1
         & 1.3 & 2.2(3)  & 4.0\\
 6.0(5)  & 1.9 & 2.8(2)  & 13.0
         & 1.8 & 2.8(3)  &  9.2
         & 1.7 & 2.6(3)  &  6.2
         & 1.6 & 2.5(3)  & 4.0\\
\hline
\end{tabular}
}
\end{table}

\clearpage

\begin{table}
\caption{Apparent half-lives of the transitions involved in the decay
of isomeric states in  $^{210}$Ra, $^{211}$Ra and $^{212}$Ra.}
\label{half-life}
\begin{tabular}{|rr|rr|rr|}
\hline
\multicolumn{2}{|c}{$^{210}$Ra} &
\multicolumn{2}{|c}{$^{211}$Ra} &
\multicolumn{2}{|c|}{$^{212}$Ra} \\
$E_\gamma$ [keV] & $T_{1/2}$ [$\mu$s] &
$E_\gamma$ [keV] & $T_{1/2}$ [$\mu$s] &
$E_\gamma$ [keV] & $T_{1/2}$ [$\mu$s] \\
\hline
 96     & 2.51(31) & 396 & 9.5(8) & 441 & 8.9(9)   \\
750     & 2.57(32) & 802 & 9.9(8) & 825 & 7.7(1.3)   \\
577     & 2.37(17) &     &        & 629 & 10.2(1.0)\\
774     & 2.34(17) &     &        & & \\
601/603 & 2.15(11) &     &        & & \\
\hline  
\end{tabular}
\end{table}

\end{document}